\documentstyle[epsfig]{aipproc}

\begin{document}

\noindent
\begin{minipage}[t]{.2\linewidth}
\leavevmode
 \hspace*{-.8cm}
\psfig{file=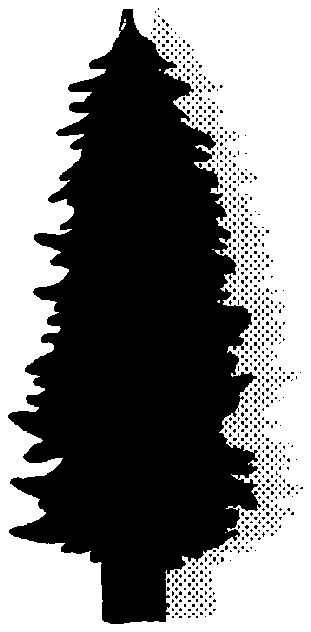,width=3cm}
\end{minipage} \hfill
\begin{minipage}[b]{.45\linewidth}
\rightline{SCIPP 01/08}
\rightline{February 2001}
\vspace{3cm}
\end{minipage}
\vskip1.5cm

\thispagestyle{empty}

\begin{center} {\large \bf Update on the Measurement of $\alpha_S$
with a 500 GeV Linear Collider}\\[2pc]
%\title{Reliability of the Models\\ and Their Match with Observations}

BRUCE A. SCHUMM and  ANDREW S. TRUITT \\
\smallskip
The University of California\\
and the\\
Santa Cruz Institute for Particle Physics\\
Santa Cruz, CA 95064
\end{center}
\vskip1cm

%\begin{abstract}
An update on the prospects for the precise measurement of
the strong coupling constant $\alpha_S$ at a high energy Linear
Collider via the three-jet rate
is presented. In particular, the issue of the distribution of
center-of-mass energies of the identified $q \bar{q}$ event sample, 
which can affect the
determination of $\alpha_S$ at the scale $Q^2=(500\rm{GeV})^2$, is addressed.
%\end{abstract}

\vfill
\begin{center}
To be published in the proceedings of the LCWS 2000
Linear Collider Workshop,
Fermilab, Batavia, IL, October 24--28, 2000
\end{center}
\bigskip
\begin{center}
{\small This
work was supported in part by the Department of Energy,
Grant \#DE-FG03-92ER40689}
\end{center}
\vfill
\clearpage
%\lefthead{LEFT head}
%\righthead{RIGHT head}
\maketitle

\setcounter{page}{1}

%
%\bigskip
%\smallskip
%\smallskip
%
%\section*{Introduction}

The idea that a 500 GeV Linear Collider could provide an
accurate measurement of the strong coupling constant
$\alpha_S$ at a scale $Q^2 = (500 \rm{GeV})^2$ has been
under consideration for some time. In Reference~\cite{schumm},
the issue of identifying a $q \bar{q}$ sample of  high enough
quality (both in terms of its purity as well as its potential
bias against $q \bar{q}$ events with hard
gluon radiation) was addressed in the context of the
determination of the three-jet rate $R_3$.
In that study, it was found that selection criteria
based on charged particle multiplicity, thrust direction,
visible energy, and momentum balance, when combined with
a b-quark antitag (to select against $t \bar{t}$ events)
and the use of electron beam polarization
(to select against $W^+W^-$ production) provided a means
of identifying a sample of
$e^+e^- \rightarrow q \bar{q}$ ($q \ne t$) at $E_{cm}~=~500$~GeV
adequate for the measurement of $\alpha_S$ to $\sim \pm 1\%$.

While the study of Reference~\cite{schumm} did include the effects
of initial state radiation and beamstrahlung on the identified
event sample, it did not address the issue of how well the
differential luminosity spectrum would have to be known in
order to achieve an accuracy of $< \pm 1\%$ on
$\alpha_S$ at $Q^2=(500\rm{GeV})^2$. That issue is addressed in
this study. A sample of simulated $e^+e^- \rightarrow q \bar{q}$
events was generated using the PYBMS~\cite{PYBMS} implementation
of PYTHIA~\cite{PYTHIA}. Jets were reconstructed according to the
$E0$~\cite{E0} jetfinding algorithm, assuming calorimetric coverage out
to $|\cos\theta| \le 0.99$. The limiting effect of calorimeter segmentation
was not considered, although it is not expected to substantially change the
conclusions of the study.

The differential luminosity spectrum of effective center-of-mass
energy of hard $e^+e^-$ collisions at the Linear Collider
can effect the $\alpha_S$ measurement in two ways. Firstly, due
to the accompanying boost, jets can coalesce when analyzed with
a fixed jet-mass resolution parameter, leading to an underestimate
of the three-jet rate $R3$. Second, the running of $\alpha_S$ itself
leads to an increase of $R3$ for events with substantial radiation
in the initial state. Both of these effects need to be understood
well enough so that their correction introduces as systematic error
substantially less than 1\%.

Figure 1 shows the ratio of the reconstructed and true three-jet
fractions $R3$ as a function of the effective center-of-mass energy
of the event sample, assuming a nominal beam energy of 250 GeV, and
a jet-mass resolution parameter of $Y_{cut} = 0.003$.
Also shown is the differential luminosity spectrum of
$e^+ e^- \rightarrow q \bar{q}$
events after the application of the event selection criteria of
Reference~\cite{schumm}.
A convolution of the two curves yields a mean value of
$$ \langle {R3_{recon} \over R3_{true}} \rangle = 0.982, $$
or about a 2\% difference.

Figure 2 shows the evolved value of $\alpha_S$ as a function of
effective center-of-mass energy, again superimposed over the
spectrum of selected events. A convolution of these two curves
yields
$$ \langle {\alpha_S^{meas} \over \alpha_S^{500}} \rangle = 1.012, $$
or about a 1\% difference.

In either case, knowledge of the differential luminosity spectrum to
about $\pm 10\%$ or so, a far looser requirement than that imposed
by other physics requirements (such as the top threshold study)
would lead to a systematic uncertainty substantially below 1\%.
Thus, it is concluded that neither of these effects presents any
problem for the accurate measurement of $\alpha_S$ at a high
energy Linear Collider.

\begin{figure}[b!] % fig 1
\centerline{
\epsfig{file=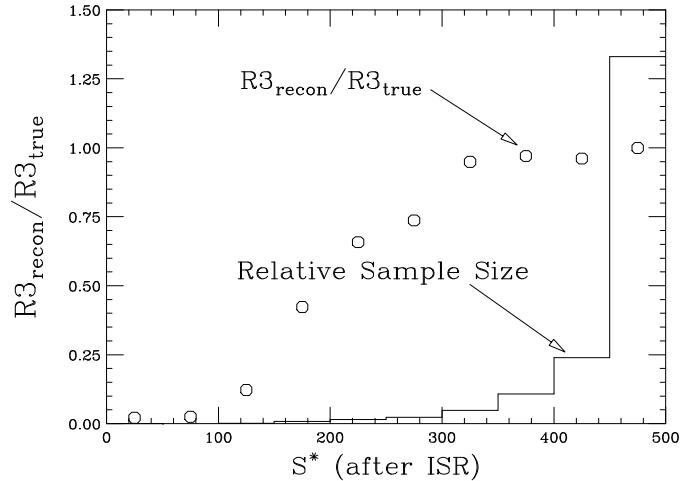,height=2.5in,width=3.5in}}
\vspace{10pt}
\caption{Measured vs. true three-jet rate $R3$ as a function 
of effective
  center-of-mass energy $S^*$ after ISR and beamstrahlung. 
Also shown is
  the $S^*$ distribution of events, again after 
ISR and beamstrahlung,
  passing the event selection requirements of Reference [1].}
\label{fig1}
\end{figure}

\begin{figure}[b!] % fig 2
\centerline{
\epsfig{file=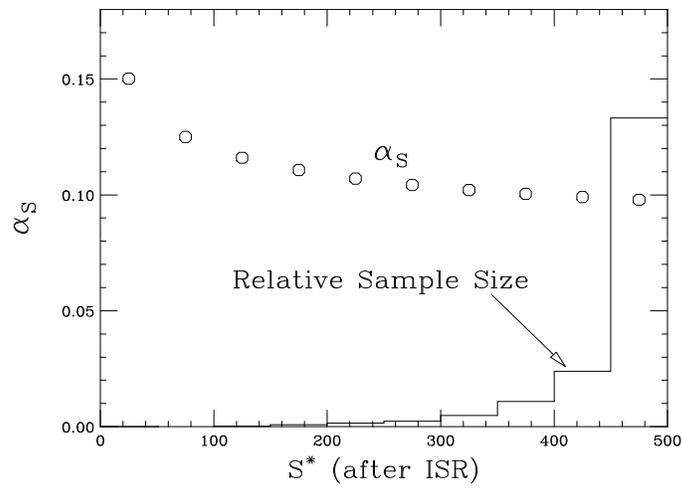,height=2.5in,width=3.5in}}
\vspace{10pt}
\caption{Evolved $\alpha_S$ as a function of effective
  center-of-mass energy $S^*$ after ISR and beamstrahlung. Also shown is
  the $S^*$ distribution of events, again after ISR and beamstrahlung,
  passing the event selection requirements of Reference [1].}
\label{fig2}
\end{figure}

\end{document}